# Exciton dynamics in atomically thin $MoS_2$: inter-excitonic interaction and broadening kinetics

Sangwan Sim,[1] Jusang Park,[1] Jeong-Gyu Song,[1] Chihun In,[1] Yun-Shik Lee,[2] Hyungjun Kim,[1,*] and Hyunyong Choi[1,†]

[1] *School of Electrical and Electronic Engineering, Yonsei University, Seoul 120-749, Korea*
[2] *Department of Physics, Oregon State University, Corvallis, OR 97331-6507, USA*

We report ultrafast pump-probe spectroscopy examining exciton dynamics in atomically thin $MoS_2$. Spectrally- and temporally-resolved measurements are performed to investigate the interaction dynamics of two important direct-gap excitons (*A* and *B*) and their associated broadening kinetics. The two excitons show strongly correlated inter-excitonic dynamic, in which the transient blue-shifted excitonic absorption originates from the internal *A-B* excitonic interaction. The observed complex spectral response is determined by the exciton collision-induced linewidth broadening; the broadening of the *B* exciton linewidth in turn lowers the peak spectral amplitude of the *A* exciton. Resonant excitation at the *B* exciton energy reveals that inter-excitonic scattering plays a more important role in determining the broadening kinetics than free-carrier scattering.

[*] E-mail: hyungjun@yonsei.ac.kr

[†] E-mail: hychoi@yonsei.ac.kr

Quasi-two-dimensional (2D) layers of thin transition metal dichalcogenides (TMDCs) have received a lot of attention in low-dimensional applications due to their excellent electronic and optical properties.[1] Molybdenum disulfide[2-10] ($MoS_2$), an interesting member of DTMC family, is an indirect band-gap semiconductor without any significant optical activity as long as it retains its bulk structural properties.[11] As the number of lattice layers decreases, however, it begins to show strong photoluminescence near the direct gap at the K point because the indirect gap near Γ point of the Brillouin zone is significantly widened by the two-dimensional quantum-confinement effect, which has a relatively weaker effect at the K point due to the high out-of-plane electron and hole masses.[11,12] Thus, in an atomically thin $MoS_2$ structure, the direct-gap exciton dynamics at the K point play a key role in determining the confinement-enhanced two-dimensional optical characteristics.

Recent work[13] suggests that the exciton dynamics[14-20] in $MoS_2$ is strongly influenced by valence band splitting, which leads to Coulomb-enhanced multi-exciton excitations, namely the *A* and *B* excitons (see the arrows in Figure 1a). Because the electron and hole (*e-h*) pair contributing to the formation of *A* and *B* excitons is closely located both in momentum and in energy space, it is expected that these two exciton states will mutually drive correlated interactions. Recent ultrafast spectroscopy research has reported some oscillatory spectra in the optical absorption responses.[21,22] However, their interpretations are focused on the single exciton dynamics, which fail to capture the nature of the interaction between the two *A* and *B* excitons. It is thus crucial to understand the internal exciton dynamics as a whole, including phenomena such as exciton absorption bleaching, exciton energy shift, and broadening of the photo-excited *e-h* pairs, to determine the dynamic origin of the two-particle *A-B* exciton interaction.[23]

Here, we report ultrafast optical pump-probe spectroscopy of multi-exciton dynamics in an ultra-thin $MoS_2$ film. By performing temporally- and spectrally-resolved measurements, we observe strong transient absorption blue-shifts in the two *A* and *B* excitons, which arises from internal pairwise repulsive excitonic interactions. The associated linewidth broadening is significant, and we find that the transient broadening is caused by exciton-exciton scattering between the pair of the *A* and *B* excitons, rather than interactions between exciton and free-carriers.

Ultrafast pump-probe spectroscopy was performed in the visible range using a 50 fs, 250 kHz Ti:sapphire laser system (Coherent RegA 9050).[24] We used two excitation wavelengths (605 and 400 nm) generated by the second harmonics of an optical parametric amplifier's (Coherent OPA) output pulses. A white-light supercontinuum created in a sapphire disk served as probe pulses (560 – 730 nm). The group-delay dispersion of white-light continuum was measured via the cross-correlation of the white-light and an 800 nm pulse on a BBO crystal.[25] By scanning the time delay for each white-light wavelength, it was possible to determine the time delay of maximum up-conversion signal. In this way, we obtained the time-profile of the white-light probe, which was used to correct the measured ultrafast dynamics for each probe wavelength.[26,27] The ultrafast dynamics was investigated by time-resolved differential transmission (DT) experiments at room temperature. In order to achieve homogeneous pump excitation, the size of pump and probe beam spots are 100 $\mu$m and 60 $\mu$m, respectively.

The $MoS_2$ film was synthesized with hot-wall furnace type system. Mo deposited (~1nm) single crystal MgO (100) substrate was placed in the center of tube furnace. The furnace was gradually heated from room temperature to 600 °C for 30 min. Then the temperature was increased from 600 °C to 1000 °C for 90 min. After keeping at 1000 °C for 10 min, the furnace was naturally cooled down to room temperature. During the reaction, Ar and $H_2S$ gas were kept flowing at a rate of 100 sccm and 5 sccm, respectively.

The Raman spectroscopy with excitation laser lines of 532 nm and atomic force microscope (AFM) were employed to identify thickness and cristallinity of $MoS_2$ film. As shown in Figure 1b, two characteristic Raman modes, $E_{2g}^1$ (381.9 cm$^{-1}$) and $A_{1g}$ (405.9 cm$^{-1}$) with the full-width-half-maximum (FWHM) of 4.9 and 3.6 cm$^{-1}$, are observed. From the peak distance (~24 cm$^{-1}$) between $E_{2g}^1$ and $A_{1g}$ modes, the layer number of our $MoS_2$ film was estimated to be 4~5.[28-30] The FWHM exhibits that the grown $MoS_2$ on MgO (100) single crystal substrate has high crystallinity comparable with the exfoliated single crystal $MoS_2$ which has the FWHM of around 3~5 cm$^{-1}$.[31] In order to confirm the thickness of sample, we transferred the $MoS_2$ film to $SiO_2$ substrate and took AFM image as shown in Figure 1c; the image analysis shows that the film is uniform with layer thickness of around 3 nm. We additionally note that the measured thickness is in agreement with the result of Raman spectroscopy of Figure 1b.[32] The size of the $MoS_2$ film after transfer is large (~ 1 cm × 1 cm).

Figures 2a and b display the spectral and temporal responses with excitation at the *B* exciton resonant energy as a function of the pump-probe delay $\Delta t$. Two positive differential transmission (DT) changes are observed at the *A* and *B* exciton energies (1.88 and 2.03 eV, respectively). Between the two exciton lines, a complex spectral response with negative DT is observed. If we use simple exciton population dynamics to interpret the pump-probe signals, we may come to an abnormal physical conclusion: if we assume that the positive DT signals at 1.88 and 2.03 eV derive from the *A* and *B* exciton population dynamics, respectively, we conclude that the higher energy exciton (*B* exciton) has a greater population than the lower energy one (*A* exciton), even for a long time $\Delta t$ of tens of ps. We note that the results do not agree with the following theoretical prediction based on a 2D Saha equation[33,34] unless the ratio of coupling constants of *A* and *B* excitons is equal to or larger than $(8.9 \times 10^{-4})^{-1} \approx 1.1 \times 10^3$.

$$N_B / N_A = (m_B^* / m_A^*) \exp(-\Delta / kT) \approx 8.9 \times 10^{-4} \ll 1, \quad (1)$$

where $N_A$ and $N_B$ are the *A* and *B* exciton populations, $m_A^* \approx 4.10 m_0$ and $m_B^* \approx 1.28 m_0$ are the *A* and *B* exciton masses, $m_0$ is the electron rest mass,[35] $\Delta \approx 0.15$ eV is the energy difference between the *A* and *B* excitons, $k$ is the Boltzmann constant and $T$ is the 300 K carrier temperature.

To resolve this issue, we have carried out more detailed spectral analyses. We fit the equilibrium absorption spectrum as a sum of two Gaussian functions; corresponding to the *A* and *B* excitons, respectively, as shown in Figure 3a.[36] Figure 3b shows the spectrally-resolved DT data, resonantly pumped at the *B* exciton energy. The DT spectra are fitted by subtracting the equilibrium absorption Gaussians (dashed line in Figure 3a) from the pump-induced absorption changes, assuming that the absorption amplitude, linewidth, and center position are not changed significantly by the photo-excitation.[22,37] We notice that the integrated areas of both the *A* and *B* absorption Gaussians show negligible changes from those of the original equilibrium Gaussian absorptions as a function of $\Delta t$ (Figure 3c). This constraint ensures that the phase-space filling is not the dominant mechanism for the observed transmission increase near 1.88 and 2.03 eV; otherwise, the pump-created excitons will reduce the integrated areas of the absorption Gaussians. As discussed later, we show that the observed transmission changes originate from the other two mechanisms, namely the spectral peak-shift and broadening of the absorption Gaussians.[36-40]

First, we discuss the transient spectral dynamics of the peak-shift in the *A* and *B* exciton energy lines. As illustrated in the inset of Figure 3d, we note that the transient blue-shift in the absorption Gaussian represents the red-shift in the DT spectra. In a strong quantum-confinement regime, the photo-generated excitons undergo many-body exciton-exciton repulsive interactions, whose typical spectroscopic signature appears as a red-shift in the DT spectra—a phenomenon known as exciton self-energy renormalization.[36,38,39,41] Figure 3d illustrates the transient peak shifts of the *A* and *B* absorption peaks (open squares and triangles, respectively) and the initial DT build-up dynamics of the *A* and *B* excitons (red and blue solid line, respectively). Although the energy shift of the *A* exciton already reaches its maximum before its DT build-up is finished, the transient energy shift of the *A* and *B* excitons shows similar dynamics. This is a classic indication of inter-excitonic interaction; in the many-body interaction regime, the energy shift arises from the mutual interactions between two exciton particles, rather than from single-particle interactions.[36,38,39,41,42] If the involved interaction comes from the *A* and *B* excitons as in our case, it is expected that the temporal dynamics of the two exciton energies should exhibit similar transients. Note that the transient energy shift of the *A* exciton differs from the build-up dynamics of the *A* exciton. Rather, the energy shift of the *A* exciton closely follow the DT build-up of the *B* exciton. This analysis suggests that pairwise internal interaction between the *A* and *B* excitons leads to the transient blue shift of their associated exciton energies.

The scenario described above is further corroborated by a comparison of the recovery dynamics of the *A* and *B* resonant energies, shown in Figure 3e. As would be expected from the two excitonic interactions, the two solid lines obtained by fitting to the transient energy shifts of the *A* and *B* excitons are nearly indistinguishable. We can faithfully fit the transient resonant energies with a single exponential fit using a time constant of about 500 fs. Recently reported ultrafast studies on $MoS_2$ suggest fast decay mechanisms such as inter-valley scattering arising from indirect band structure and surface-trap mediated scattering.[21,43] We note that the former investigations may not be applied to our study, because the lack of *2H* stacking in the CVD grown few-layer $MoS_2$ could lead to negligible interlayer coupling between the layers.[9] Our investigation suggests that the exciton-exciton scattering, revealed as blue-shift dynamics on the 500 fs time scale, is one important factor for understanding the fast dynamics in addition to the surface-trap mediated scattering. Note that the exciton ionization is not responsible for the fast decay;[33] although the calculated exciton binding energies of monolayer and bilayer MoS$_2$ are 0.9 eV and 0.4 eV,[44] respectively, a recent ultrafast study[21] on the 2D MoS$_2$ thin films shows that the exciton dynamics of monolayer and a few layer MoS$_2$ reveals no significant difference at room temperature.

The second mechanism for explaining the DT imbalance (Figure 2) between the *A* and *B* excitons is the pump-induced exciton linewidth $\Gamma$ broadening.[21,36,40] In a simple excitonic absorption picture, increased exciton density due to pump excitation leads to decreased time between exciton collisions, $\Delta\tau$, or an increased spectral $\Gamma$.[36] Figure 4a illustrates the way in which exciton broadening contributes to the observed complex DT spectra. We emphasize that the spectral response of the *A* exciton is determined not only by the spectral amplitude of the *A* exciton itself, but is also affected by the broadened negative DT shoulder of the *B* exciton, as illustrated in Figure 4b. Thus, we understand that the spectrally-broadened $\Gamma$ of the two excitons simultaneously influence the DT imbalance, which naturally explains why the spectral amplitude of the *A* exciton energy is smaller than that of the *B* exciton (black downward arrow in Figure 4b). Figure 4c shows the temporal dynamics of the $\Gamma$ broadening fitted by two exponential functions. The fast decay components are 460 and 530 fs for the *A* and *B* excitons, respectively. Whereas the initial decay exhibits a feature similar to the transient blue shift of the two exciton energies (Figure 3e), the transient $\Gamma$ broadening shows an additional decay component of about 400 ps, which may be attributed to long-lived exciton radiative transition.[13]

The physical origin of the $\Gamma$ broadening is further clarified by comparing the spectral response of excitation at the *B* exciton resonant energy with that of excitation at a higher energy (3.01 eV). The 3.01 eV excitation, the so-called *C* exciton transition, is known to cause an inter-band transition from the deep valence band to the conduction band.[45] Because the pump-created holes in the deep valence band have a low probability of scattering to the K point, a relatively small number of carriers contributes to the *A* and *B* exciton populations; in other words, the resonant *C* exciton excitation creates less exciton density. Figure 4d compares the DT spectrum for the excitation at the *C* exciton resonance (3.01 eV, green filled circles) with that for the *B* exciton resonance (1.88 eV, black open circles), keeping the photo-excited carrier densities constant ($9 \times 10^{10}$ cm$^{-2}$ per layer). Resonant excitation at the *C* exciton shows a similar DT spectrum to that of the *B* exciton, but with much larger amplitude. If we assume the integrated areas of the DT spectra are the same in both cases, it implies that the $\Gamma$ suffers more broadening for the resonant excitation at the *C* exciton. This result is expected because the $\Gamma$ broadening exhibits a much broader feature (and thus more oscillatory DT spectra) when the excitons collide with free-carriers than the case of collision with other excitons.[36,40] In other words, the narrow $\Gamma$ for the resonant excitation at the *B* exciton is due to a pairwise inter-excitonic collision rather than scattering between exciton and free-carriers. Therefore, we expect that the red-shift in the DT spectra for the resonant *C* excitation will be small. Indeed, this is exactly what we observe in Figure 3d, where a weak red-shift in the DT spectra is observed for the excitation at the *C* exciton resonant energy. This result also supports the idea that the transient blue-shift of the excitonic absorption is due to an internal excitonic interaction.

Finally, the pump-fluence dependent study provides more information on the blue-shift and broadening of excitons. Figure 5a shows the observed DT spectra for different pump fluences with a fixed pump-probe time delay of 0.1 ps. We note that excitations with different pump fluences lead to similar spectral shape, which suggests that the origin of spectral response does not depends on the exciton population. As shown in Figure 5b and 5c, both the blue-shift and broadening of *A* and *B* excitons reveal linear pump-fluence dependence. This characteristic is in accordance with our analysis of the exciton-exciton interactions, i.e. as more excitons are injected by photoexcitation, more excitonic interactions take place.[36]

To conclude, our ultrafast measurements on exciton dynamics reveals the following two main aspects. First, the *A* and *B* excitons located at the K point repulsively interact with each other, leading to a similar temporal evolution of the transient blue-shift of their absorption Gaussians. The peak energy-shift and the DT build-up transients are explained by these *A*-*B* inter-excitonic interactions. Second, the DT imbalance between the *A* and *B* excitons is explained by pump-induced $\Gamma$ broadening. The resonant excitation at the *C* exciton further confirms the inter-excitonic scattering.

The work at Yonsei was supported by Basic Research Program through the National Research Foundation of Korea (NRF) funded by the Ministry of Education, Science and Technology (No. 2011-0013255), the NRF grant funded by the Korean government (MEST) (NRF-2011-220-D00052, No.2011-0028594, No.2011-0032019), the Converging Research Center Program through the Ministry of Education, Science and Technology (No. 2013K000173) and the LG Display academic industrial cooperation program. The work at Oregon State University was supported by the National Science Foundation (DMR-1063632).

**References**

[1] Q. H. Wang, K. Kalantar-Zadeh, A. Kis, J. N. Coleman, M. S. Strano, Nat. Nanotechnol. **7,** 699 (2012).

[2] M. S. Choi, G. H. Lee, Y. J. Yu, D. Y. Lee, S. H. Lee, P. Kim, J. Hone, W. J. Yoo, Nat. Commun. **4,** 1624 (2013).

[3] B. Radisavljevic, A. Kis, Nat. Nanotechnol. **8,** 147 (2013).

[4] S. Helveg, J. V. Lauritsen, E. Laegsgaard, I. Stensgaard, J. K. Norskov, B. S. Clausen, H. Topsoe, F. Besenbacher, Phys. Rev. Lett. **84,** 951 (2000).

[5] C. Rice, R. J. Young, R. Zan, U. Bangert, D. Wolverson, T. Georgiou, R. Jalil, K. S. Novoselov, Phys. Rev. B **87,** 081307 (2013).

[6] H. Li, G. Lu, Z. Y. Yin, Q. Y. He, H. Li, Q. Zhang, H. Zhang, Small **8,** 682 (2012).

[7] Y. Feldman, E. Wasserman, D. J. Srolovitz. R. Tenne, Science **267,** 222 (1995).

[8] R. Wang, B. A. Ruzicka, N. Kumar, M. Z. Bellus, H. Y. Chiu, H. Zhao, Phys. Rev. B **86,** 045406 (2012).

[9] N. Kumar, S. Najmaei, Q. N. Cui, F. Ceballos, P. M. Ajayan, J. Lou, H. Zhao, Phys. Rev. B **87,** 161403 (2013).

[10] L. M. Malard, T. V. Alencar, A. P. M. Barboza, K. F. Mak, A. M. de Paula, Phys. Rev. B **87,** 201401 (2013).

[11] K. F. Mak, C. Lee, J. Hone, J. Shan, T. F. Heinz, Phys. Rev. Lett. **105,** 136805 (2010).

[12] A. Splendiani, L. Sun, Y. B. Zhang, T. S. Li, J. Kim, C. Y. Chim, G. Galli, F. Wang, Nano Lett. **10,** 1271 (2010).

[13] T. Korn, S. Heydrich, M. Hirmer, J. Schmutzler, C. Schuller, Appl. Phys. Lett. **99,** 102109 (2011).

[14] T. Elsaesser, J. Shah, L. Rota, P. Lugli, Phys. Rev. Lett. **66,** 1757 (1991).

[15] T. Elsaesser, R. J. Bauerle, W. Kaiser, H. Lobentanzer, W. Stolz, K. Ploog, Apll. Phys. Lett. **54,** 256 (1989).

[16] T. Shih, K. Reimann, M. Woerner, T. Elsaesser, I. Waldmuller, A. Knorr, R. Hey, K. H. Ploog, Phys. Rev. B **72,** 195338 (2005).

[17] C. Y. Sung, T. B. Norris, A. Afzali-Kushaa, G. I. Haddad, Appl. Phys. Lett. **68,** 435 (1996).

[18] T. S. Sosnowski, T. B. Norris, H. Jiang, J. Singh, K. Kamath, P. Bhattacharya, P. Phys. Rev. B **57,** R9423 (1998).

[19] J. Urayama, T. B. Norris, J. Singh, P. Bhattacharya, Phys. Rev. Lett. **86,** 4930 (2001).

[20] W. Sha, T. B. Norris, W. J. Schaff, K. E. Meyer, Phys. Rev. Lett. **67,** 2553 (1991).

[21] H. Shi, R. Yan, S. Bertolazzi, J. Brivio, B. Gao, A. Kis, D. Jena, H. G. Xing, L. Huang, ACS Nano **7,** 1072 (2013).

[22] R. Wang, B. A. Ruzicka, N. Kumar, M. Z. Bellus, H. Y. Chiu, H. Zhao, arXiv:1110.6643 (2011).

[23] J. Kono, S. T. Lee, M. S. Salib, G. S. Herold, A. Petrou, B. D. McCombe, Phys. Rev. B **52,** R8654 (1995).

[24] T. B. Norris, Opt. Lett. 17, 1009 (1992).

[25] J. Urayama, Ph.D. thesis, The University of Michigan, 2002.

[26] A. Maciejewski, R. Naskrecki, M. Lorenc, M. Ziolek, J. Karolczak, J. Kubicki, M. Matysiak, M. Szymanski, J. Mol. Struct. **555,** 1 (2000).

[27] T. Nakayama, Y. Amijima, K. Ibuki, K. Hamanoue, Rev. Sci. Instrum. **68,** 4364 (1997).

[28] S.-L. Li, H. Miyazaki, H. Song, H. Kuramochi, S. Nakaharai, K. Tsukagoshi, ACS Nano **6,** 7381 (2012).

[29] H. Li, Q. Zhang, C. C. R. Yap, B. K. Tay, T. H. T. Edwin, A. Olivier, D. Baillargeat, Adv. Funct. Mater. **22,** 1385 (2012).

[30] C. Lee, H. Yan, L. E. Brus, T. F. Heinz, J. Hone, S. Ryu, ACS Nano **4,** 2695 (2010).

[31] K.-K. Liu, W. Zhang, Y.–H. Lee, Y.–C. Lin, M.–T. Chang, C.–Y. Su, C.–S. Chang, H. Li, Y. Shi, H. Zhang, C.–S. Lai, L. -J. Li, Nano Lett. **12,** 1538 (2012).

[32] B. Radisavljevic, A. Radenovic, J. Brivio, V. Giacometti, A. Kis, A. Nat. Nanotechnol. **6,** 147 (2011).

[33] D. S. Chemla, D. A. B. Miller, P. W. Smith, A. C. Gossard, W. Wiegmann, IEEE J. Quant. Electron. **QE-20,** 265 (1984).

[34] R. A. Kaindl, D. Hagele, M. A. Carnahan, D. S. Chemla, Phys. Rev. B **79,** 045320 (2009).

[35] B. Visic, R. Dominko, M. K. Gunde, N. Hauptman, S. D. Skapin, M. Remskar, Nanoscale Res. Lett. **6,** 593 (2011).

[36] D. R. Wake, H. W. Yoon, J. P. Wolfe, H. Morkoc, Phys. Rev. B **46,** 13452 (1992).

[37] G. N. Ostojic, S. Zaric, J. Kono, V. C. Moore, R, H. Hauge, R. E. Smalley, Phys. Rev. Lett. **94,** 097401 (2005).

[38] N.Peyghambarian, H. M. Gibbs, J. L. Jewell, A. Antonetti, A. Migus, D. Hulin, A. Mysyrowicz, Phys. Rev. Lett. **53,** 2433 (1984).

[39] F. Vouilloz, D. Y. Oberli, F. Lelarge, B. Dwir, E. Kapon, Solid State Commun. **108,** 945 (1998).

[40] A. Honold, L. Schultheis, J. Kuhl, C. W. Tu, Phys. Rev. B **40,** 6442 (1989).

[41] D. Hulin, A. Mysyrowicz, A. Antonetti, A. Migus, W. T. Masselink, H. Morkoc, H. M. Gibbs, N. Peyghambarian, Phys. Rev. B **33,** 4389 (1986).

[42] S. Das Sarma, D. W. Wang, Phys. Rev. Lett. **84,** 2010 (2000).

[43] N. Kumar, J. Q. He, D. W. He, Y. S. Wang, H. Zhao, J. Appl. Phys. **113,** 133702 (2013).

[44] T. Cheiwchanchamnangij, W. R. L. Lambrecht, Phys. Rev. B **85,** 205302 (2012).

[45] J. P. Wilcoxon, G. A. Samara, Phys. Rev. B **51,** 7299 (1995).

**Figure captions**

FIG. 1. (color online) (a) Simplified band structure of bulk $MoS_2$. The black line is the lowest conduction band and two orange lines are the highest valence bands split by the interlayer interactions. The two arrows are the direct-gap exciton transitions, namely the *A* (red arrow) and *B* (blue arrow) excitons. (b) Raman spectrum with 532 nm excitation laser pulse of thin $MoS_2$ film sample. (c) AFM image for thin $MoS_2$ film sample. Inset: the height profile of AFM image in main panel.

FIG. 2. (color online) Spectrally- and temporally-resolved ultrafast response after resonant excitation at the *B* exciton with 10 $\mu J/cm^2$ of pump fluence. (a) The DT spectra in the photon-energy range between 1.7 and 2.15 eV and (b) the corresponding temporal dynamics with five different probe-photon energies are shown. Inset: a schematic band structure of $MoS_2$ near the K point of the Brillouin zone.

FIG. 3. (color online) Ultrafast inter-excitonic transients. (a) Absorption spectrum (thick gray line) and the corresponding fit (dashed black) using the sum of two *A* (red) and *B* (blue) absorption Gaussians are shown. (b) Spectrally-resolved DT spectra (black open circles) with excitation at the *B* exciton resonance are shown as a function of $\Delta t$. The red and blue dashed lines indicate the equilibrium peak position of the *A* and *B* excitons, respectively. The small arrows indicate the red-shifts in the DT spectra. (c) Spectrally-integrated DT areas of the absorption Gaussians. (d) Initial build-up dynamics of the *A* and *B* excitons (red and blue lines, respectively) and the transient blue-shifts of the *A* and *B* exciton energies (open squares and triangles, respectively). Inset: a schematic illustrating the transient blue-shift of the absorption Gaussians ($\alpha_0$: without the pump, $\alpha_{shifted}$: with the pump) leads to the red-shift of the DT spectra of $\Delta T$ (black solid line). (e) Recovery dynamics of the transient blue-shift and the corresponding single-exponential fit are shown for the *A* and *B* exciton energies (red and blue lines, respectively).

FIG. 4. (color online) Exciton linewidth broadening. (a) A schematic illustration for the exciton $\Gamma$ broadening. The gray line shows the equilibrium absorption without the pump. The photo-induced collision leads to the broadening of the *A* and *B* excitons (red and blue lines, respectively). (b) The corresponding DT spectra for the *A* ($\Delta T_A$, dashed red line) and the ($\Delta T_B$, dashed blue line) excitons are schematically shown. The overall DT spectra ($\Delta T = \Delta T_A + \Delta T_B$, black solid line) show the DT amplitude at the *A* exciton is smaller than that of the $\Delta T_A$, as indicated by the black downward-facing arrow. (c) Temporal dynamics of the $\Gamma$ broadening for the *A* and *B* excitons (red open squares and blue open triangles, respectively) and the corresponding bi-exponential fits (red solid for the *A* exciton and blue solid for the *B* exciton) after resonant excitation at the *B* exciton energy. (d) DT spectra with resonant excitation at the *B* exciton energy (red open circles) and those of at the *C* exciton (green filled circles) are shown. The solid lines (green for the resonant excitation at the *C* exciton and black for that at the *B* exciton) represent the corresponding fits to the measured DT spectra at $\Delta t = 0.1$ ps. The red and blue dashed lines indicate the equilibrium peak positions of the *A* and *B* exciton energies, respectively.

FIG. 5. (color online) Pump-fluence dependent spectral responses with resonant excitation at the *B* exciton. (a) DT spectra with different pump fluences of $30\ \mu J/cm^2$, $20\ \mu J/cm^2$ and 10

$\mu$J/cm$^2$ at a fixed pump-probe time delay of 0.1 ps. (b) Pump-fluence dependence of blue-shift of the *A* and *B* exciton center. (c) Linewidth broadening of the *A* and *B* excitons as a function of pump-fluence. In (b) and (c), red open squares and blue open triangles correspond to the *A* and *B* excitons, respectively. Gray solid line is an eye-guide for a linear response.

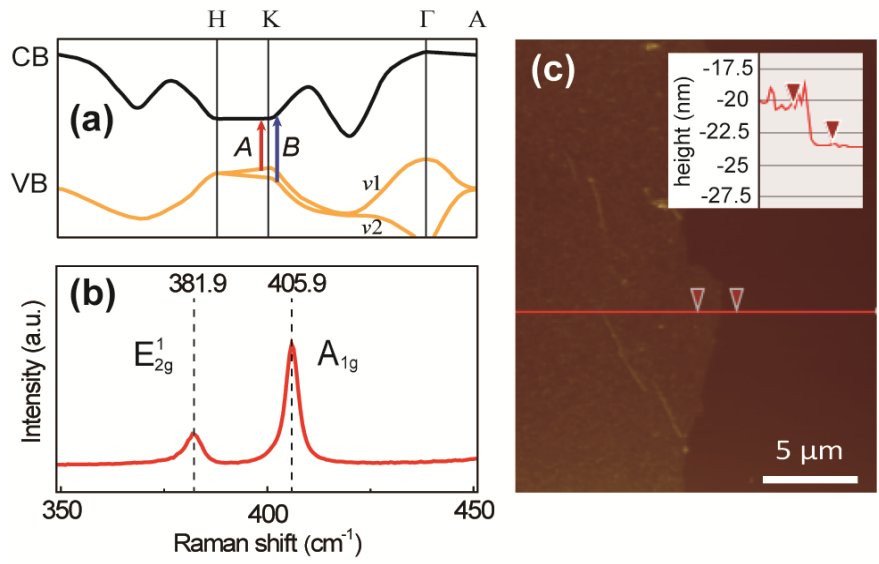

FIG. 1 S. Sim *et al.*,

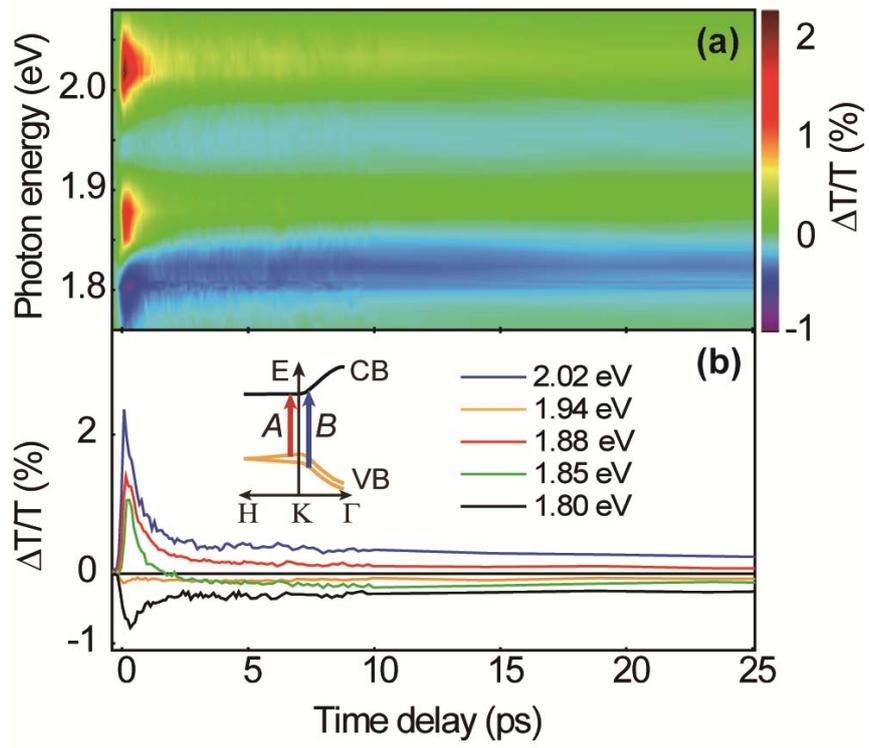

FIG. 2 S. Sim *et al.*,

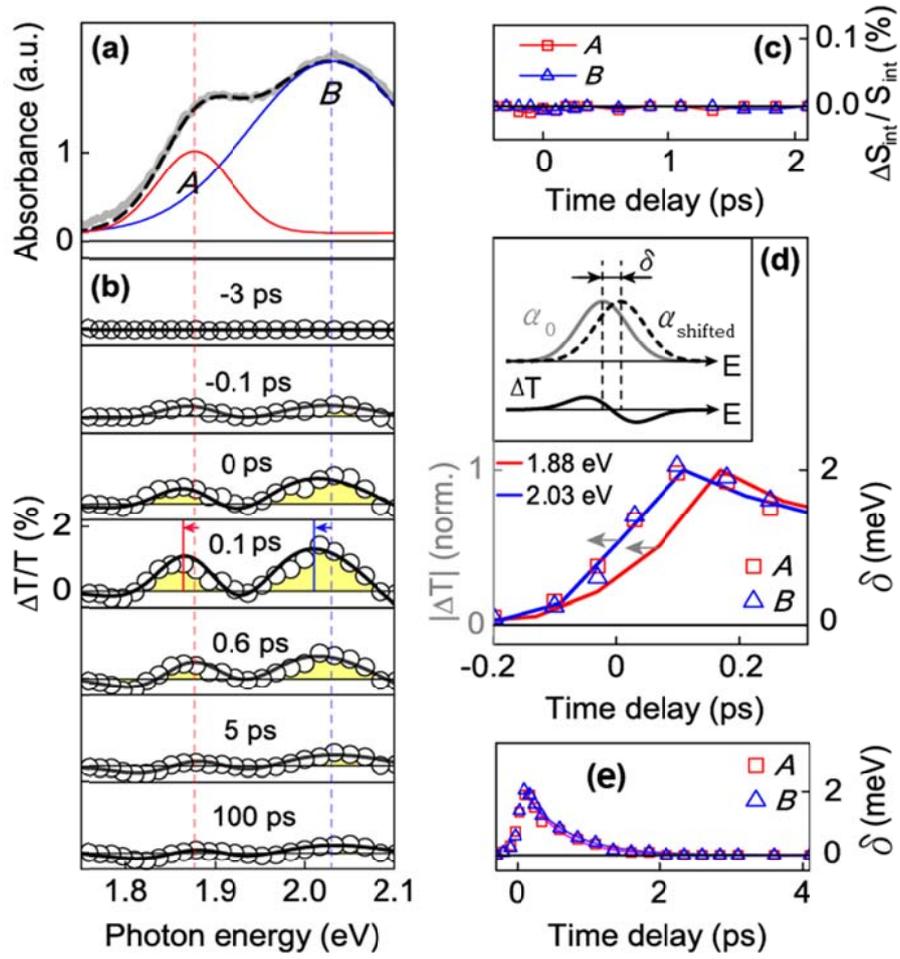

FIG. 3 S. Sim *et al.*,

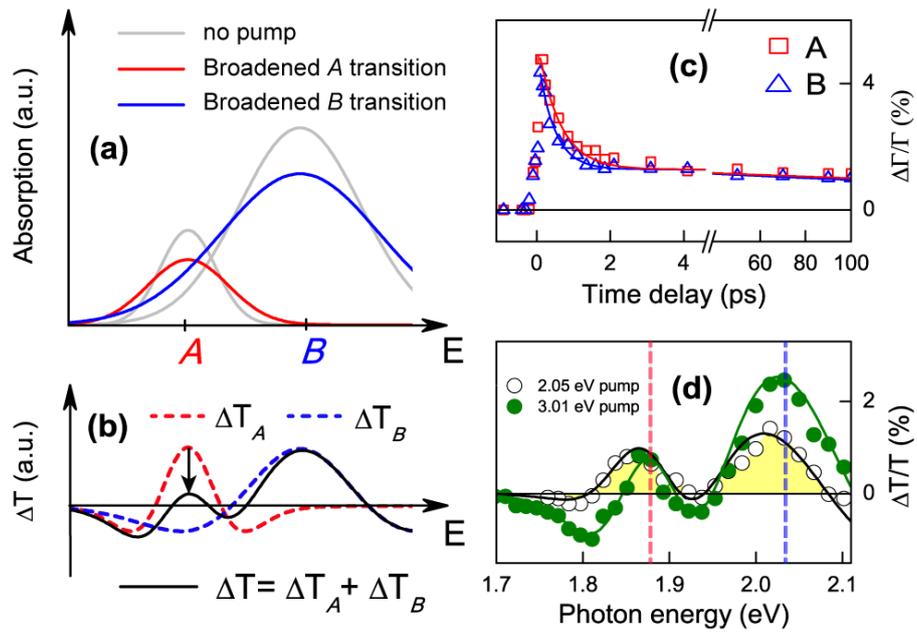

FIG. 4 S. Sim *et al*.,

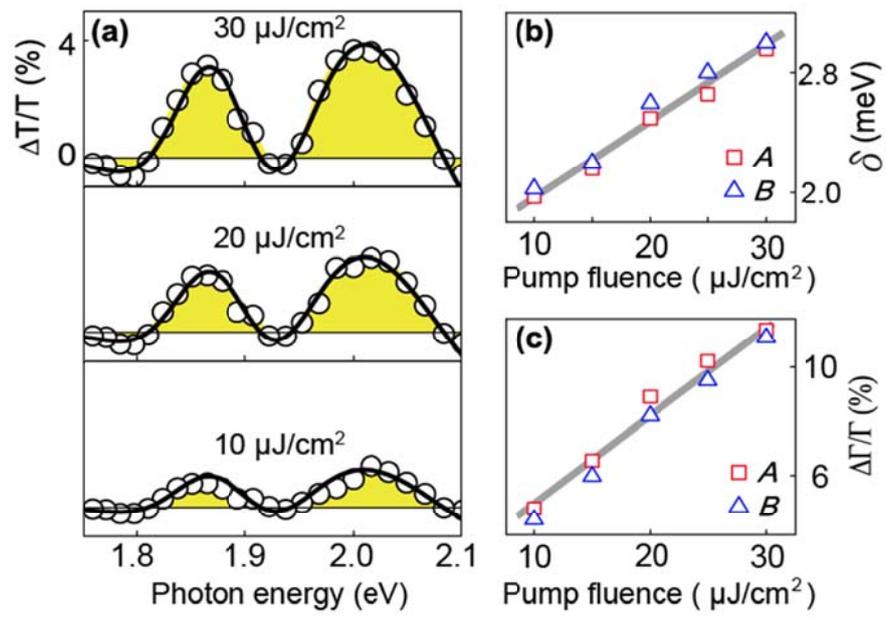

FIG. 5 S. Sim *et al.*,